\documentclass[runningheads]{llncs}
\usepackage{graphicx,amsmath,amssymb,mathtools,url,lineno}
\newcommand{\R}{\mathbb R}
\newcommand{\Z}{\mathbb Z}
\newcommand{\al}{\alpha}
\newcommand{\be}{\beta}

\newcommand{\de}{\delta}
\newcommand{\De}{\Delta}
\newcommand{\la}{\lambda}
\newcommand{\La}{\Lambda}

\newcommand{\ep}{\varepsilon}

\newcommand{\std}{\mathrm{std}}
\newcommand{\RMSE}{\mathrm{RMSE}}
\newcommand{\MAE}{\mathrm{MAE}}
\newcommand{\MAPE}{\mathrm{MAPE}}
\newcommand{\AMD}{\mathrm{AMD}}

\newcommand{\vol}{\mathrm{Vol}}

\newcommand{\id}{\mathrm{id}}

\newcommand{\bs}{\hfill $\blacksquare$}

\newcommand{\vl}{\,:\,}

\begin{document}
\title{Fast predictions of lattice energies by continuous isometry invariants of crystal structures\thanks{Supported by £3.5M EPSRC grant `Application-driven Topological Data Analysis'.}}
\titlerunning{Fast predictions of lattice energies by continuous isometry invariants}
%
\author{Jakob Ropers\inst{1}\orcidID{} \and
Marco M Mosca\inst{1}\orcidID{} \and
Olga Anosova\inst{1}\orcidID{0000-0003-4134-4398} \and
Vitaliy Kurlin\inst{1}\orcidID{0000-0001-5328-5351} \and
Andrew I Cooper\inst{1}\orcidID{}
}
\authorrunning{J. Ropers et al.}
%
\institute{University of Liverpool, Liverpool L69 3BX, UK
\email{vkurlin@liv.ac.uk}\\
\url{http://kurlin.org}}
\maketitle              
\begin{abstract}
Crystal Structure Prediction (CSP) aims to discover solid crystalline materials by optimizing periodic arrangements of atoms, ions or molecules.
CSP takes weeks of supercomputer time because of slow energy minimizations for millions of simulated crystals.
The lattice energy is a key physical property, which determines  thermodynamic stability of a crystal but has no simple analytic expression.
Past machine learning approaches to predict the lattice energy used slow crystal descriptors depending on manually chosen parameters.
The new area of Periodic Geometry offers much faster isometry invariants  that are also continuous under perturbations of atoms.
Our experiments on simulated crystals confirm that a small distance between the new invariants guarantees a small difference of energies.
We compare several kernel methods for invariant-based predictions of energy and achieve the mean absolute error of less than 5kJ/mole or 0.05eV/atom on a dataset of 5679 crystals.  

\end{abstract}

\section{Motivations, problem statement and overview of results}
\label{sec:intro}

Solid crystalline materials (\emph{crystals}) underpin key technological advances from solid-state batteries to therapeutic drugs.
Crystals are still discovered by trial and error in a lab, because their properties are not yet expressed in terms of crystal geometries.
This paper makes an important step towards understanding the structure-property relations, for example how an energy of a crystal depends on its geometric structure.
The proposed methods belong to the recently established area of Periodic Geometry, which studies geometric descriptors (\emph{continuous isometry invariants}) and metrics on a space of all periodic crystals.
\medskip

The most important property of a crystal is the energy of its crystal structure, which is usually called the \emph{lattice energy} or \emph{potential energy surface} or 
\emph{energy landscape} \cite{wales2018exploring}.
This lattice energy hints at thermodynamic stability of a crystal, whether such a crystal can be accessible for synthesis in a lab and can remain stable under application conditions. 
Since the lattice energy has no closed analytic expression, calculations are always approximate, from the \emph{force field} (FF) level \cite{niketic2012consistent} to the more exact density functional theory (DFT) \cite{gross2013density}.  
\medskip

\begin{figure}
\includegraphics[height=40mm]{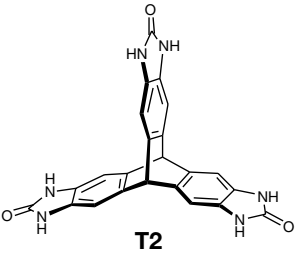}
\includegraphics[height=40mm]{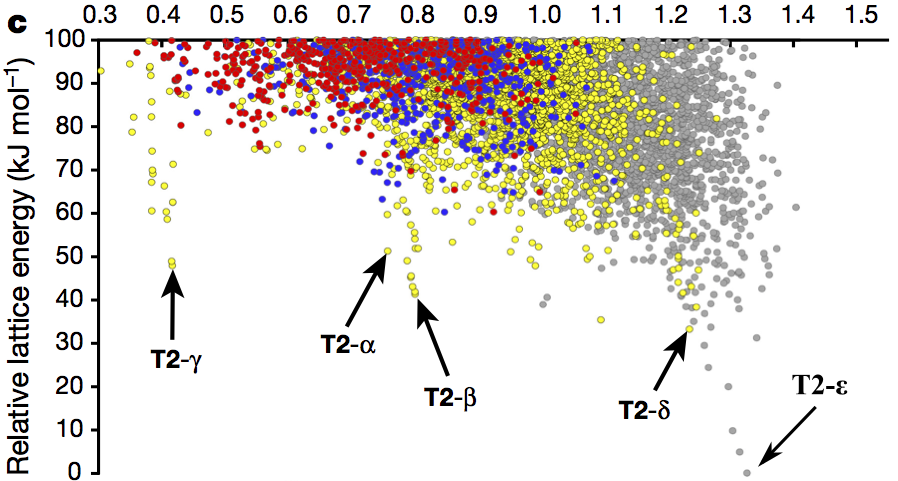}
\caption{\textbf{Left}: T2 molecule. \textbf{Right}: energy-vs-density plot of 5679 predicted crystals, five polymorphs were synthesized, most recent T2-$\ep$ crystal is added to \cite[Fig.~2d]{pulido2017functional}.} 
\label{fig:T2CSP}
\end{figure}

Our experiments use the lattice energy obtained by force fields for the CSP data of 5679 nanoporous T2 crystals in Fig.~\ref{fig:T2CSP}predicted by our colleagues \cite{pulido2017functional}.

Traditionally a periodic crystal is stored in a Crystallographic Information File (CIF).
This file specifies a linear basis $v_1,v_2,v_3$ of $\R^3$, which spans the \emph{unit cell} $U=\{\sum\limits_{i=1}^3 c_i v_i \mid 0\leq c_i<1\}$, generates the \emph{lattice} $\La=\{\sum\limits_{i=1}^3 c_i v_i \mid c_i\in\Z\}$.
Then a crystal can be obtained as the infinite union of lattice translates $M+\La=\{p+v \vl p\in M, v\in\La\}$ from a finite set (\emph{motif}) of points $M\subset U$ in the cell $U$.
The representation $M+\La$ is simple but is highly ambiguous in the sense that infinitely many pairs (cell, motif) generate equivalent crystals, see \cite[Fig.~2]{anosova2021isometry}.
\medskip

The main novelty of our approach to energy predictions is using the fast computable and easily interpretable invariants of crystals.
The concept of an invariant has a rigorous definition after we fix an equivalence relation on objects in question.
Since crystal structures are determined in a rigid form, the most natural equivalence is rigid motion or \emph{isometry}, which is any map that preserves interpoint distances, for example a composition of translations and rotations in Euclidean space $\R^3$.
Any orientation-preserving isometry can be realized as a \emph{rigid motion}, which is a continuous family $f_t$, $t\in[0,1]$, of isometries starting from the identity map $f_0=\id$.
Since any general isometry is a composition of a single reflection and a rigid motion, we consider isometry as our main \emph{equivalence relation} on crystals.
Later we can also take into account a sign of orientation. 
\medskip

An \emph{isometry invariant} $I$ is a crystal property or a function, say from crystals to numbers, preserved by isometry.
So if crystals $S,Q$ are isometric then $I(S)=I(Q)$.
The classical example invariants of a crystal $S$ are the symmetry group (the group of isometries that map $S$ to itself) and the volume of a minimal (\emph{primitive}) unit cell.
Example non-invariants are unit cell parameters (edge-lengths and angles) and fractional coordinates of atoms in a cell basis.
\medskip

Many widely used isometry invariants including symmetry groups break down (are \emph{discontinuous}) under perturbations of atoms, which always exist in real crystals at finite temperature.
Perturbations are also important for distinguishing simulated crystals obtained via Crystal Structure Prediction (CSP).
Indeed, CSP iteratively minimizes the lattice energy and inevitably stops at some approximation to a local minimum \cite{oganov2011modern}.
Hence, after many random initializations, we likely get many near duplicate structures around the same local minumum.
\medskip

\begin{figure}
\includegraphics[height=36mm]{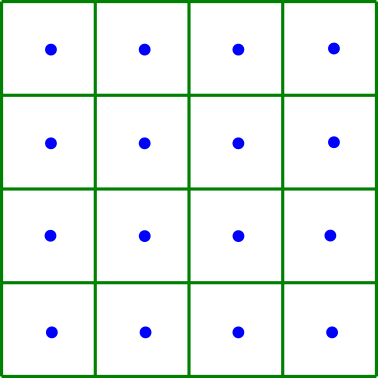}
\hspace*{4mm}
\includegraphics[height=36mm]{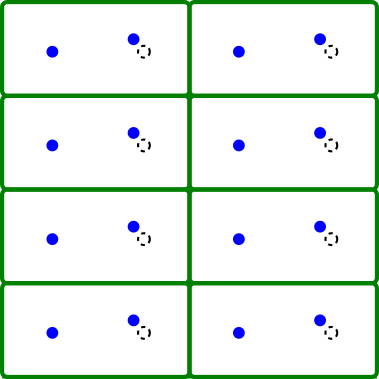}
\hspace*{4mm}
\includegraphics[height=36mm]{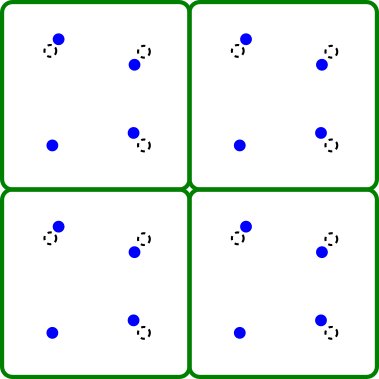}
\caption{
Most past invariants are discontinuous under perturbations above, for example symmetry groups and sizes of primitive or reduced cells. 
Recent isometry invariants \cite{mosca2020voronoi,widdowson2022average,widdowson2021pointwise} continuously quantify similarities between perturbed periodic structures.
} 
\label{fig:perturbations}
\end{figure}

Since any perturbation of points keeping their periodicity (but not necessarily an original unit cell) produces a new close structure, all periodic structures form a continuous space.
Then any CSP dataset can be viewed as a discrete sample from the underlying continuous space of periodic structures.
The lattice energy is a function on this crystal space whose geometry needs to be understood.
The problem below is a key step towards describing structure-property relations.
\medskip

\noindent
\textbf{Properties-from-invariants problem}.
Find suitable isometry invariants that justifiably predict desired properties of crystals such as the lattice energy.
\bs
\medskip
 
The proposed invariants to tackle the above problem are average minimum distances (AMD) \cite{widdowson2022average}.
AMD is an infinite sequence of isometry invariants whose values change by at most $2\ep$ if given points are perturbed in their $\ep$-neighborhoods.
A thousand of AMD invariants can be computed in milliseconds on a modest desktop for crystals with hundreds of atoms in a unit cell \cite[appendix~D]{widdowson2022average}.
\medskip
 
The above continuity of AMD guarantees that perturbed crystals have close AMD values.
Then such a theoretically continuous invariant can be tested for checking continuity of energy under crystal perturbations.
The first contribution is an experimental detection of constants $\la$ and $\de$ such that, for any smaller distance $d<\de$ between AMD vectors, the corresponding crystals have a lattice energy difference within $\la d$, usually within 2kJ/mole.
Past invariants have no such a constant to quantify continuity of energy in this way.
For example, close values of density, RMSD \cite{chisholm2005compack}, PXRD \cite{sacchi2020same} don't guarantee close values of energy.
\medskip

The second contribution is the demonstration that several kernel methods can achieve a mean absolute error of less than 5kJ/mole by using only isometry invariants without any chemical data.
The key achievement is the time of less than 10 min for training by using a modest desktop on a dataset of 5679 structures, while energy predictions take milliseconds per crystal on average.   
\medskip
 
Section~\ref{sec:review} reviews closely related past work using crystal descriptors for machine learning of the lattice energy.
Section~\ref{sec:geometry} reminds the recently introduced isometry invariants of periodic point sets and their properties.
Section~\ref{sec:continuous_energy} quantifies continuity of energy in terms of AMD invariants.
Section~\ref{sec:energy_predictions} describes how the energy of a crystal can be predicted from its AMD invariants by using several kernel methods.
Section~\ref{sec:discussion} discusses limitations and potential developments.

\section{Review of related machine learning approaches}
\label{sec:review}

This section reviews the closest related work about energy predictions for infinite periodic crystals.
The same problem is simpler for a single molecule \cite{smith17ani}.
\medskip

Energy predictions use various representations of crystals.
We review only geometric descriptors that are closest to isometry invariants in section~\ref{sec:geometry}. 
\medskip

The partial radial distribution function (PRDF) is based on the density of atoms of type $\be$ in a shell of radius $r$ and width $dr$ centered
around an atom of type $\al$ \cite{schutt14represent}.
Since atom types are essentially used, the PRDF can be best for comparing crystals that are composed of the same atom types.
Due to averaging across all atoms of a type $\al$ within a unit cell, the PRDF is independent of a cell choice.
A similar distance-based fingerprint was introduced earlier by Valle and Oganov \cite{valle2010crystal}.
Since only pairwise distances are used, these descriptors are isometry invariants and likely continuous under perturbations shown in Fig.~\ref{fig:perturbations}.
\medskip

Completeness or uniqueness of a crystal with a given PRDF is unclear yet, but can be theoretically possible for a large enough radius $r$.
Practical computations require choices or the distance thresholds $r$ and $dr$, which can affect the PRDF.
Schutt et al. confirm in \cite[Table~I]{schutt14represent} that the PRDF outperforms non-invariant features such as the Bravais matrix of cell parameters.
The mean absolute error (MAE) of energy predictions based on PRDF is 0.68eV/atom or 65.6kJ/mole.
\medskip

Another way to build geometric attributes of a crystal structure is to use Wigner-Seitz cells (also called Dirichlet or Voronoi domains) of atoms.
Ward et al. \cite{ward17including} used 271 cell-based geometric and chemical attributes to achieve the MAE of 0.09eV/atom or 8.7kJ/mole for predicting the formation enthalpy.
\medskip 

An extensible neural network potential \cite[Fig.~4]{smith17ani} has further improved the mean absolute error (MAE) to 1.8kcal/mole=7.56kJ/mole. 
The most advanced approach by Egorova et al. \cite{egorova2020multifidelity} predicts the difference between the accurate DFT energy and its force field approximation a with MAE less than 2kJ/mole by using GGA DFT (PBE) calculations and symmetry function descriptors \cite{behler2011atom}.

\section{Key definitions and recent results of Periodic Geometry}
\label{sec:geometry}

This section reviews more recent work in the new area of Periodic Geometry \cite{anosova2021introduction}, which studies the metric geometry on the space of all periodic structures.
Nuclei of atoms are better defined physical objects than chemical bonds, which depend on many thresholds for distances and angles.
Hence the most fundamental model of a crystal is a periodic set of zero-sized points representing all atomic centers.
\medskip

Though chemical elements and other physical properties can be easily added to invariants as labels of points, the experiments in \cite{edels2021,widdowson2020asymptotic} and sections~\ref{sec:continuous_energy}, \ref{sec:energy_predictions} show that the new invariants can be enough to infer some chemistry from geometry.
\medskip

\begin{figure}
\includegraphics[width=\textwidth]{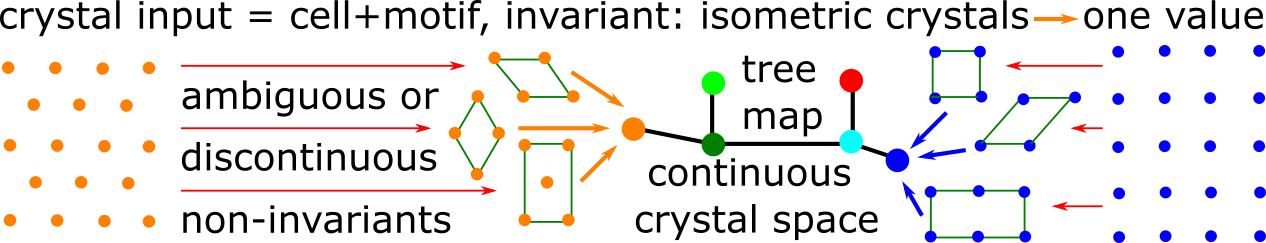}
\caption{
Any periodic sets, for example the hexagonal and square lattices, can be represented by infinitely many pairs (cell, motif).
This ambiguity can be resolved only by a complete invariant that should continuously parameterize the crystal space.
} 
\label{fig:crystal_invariants}
\end{figure}

The symbol $\R^n$ denotes Euclidean space with \emph{Euclidean} distance $|p-q|$ between points $p,q\in\R^n$.
Motivated by a traditional representation of a crystal by a Crystallographic Information File, a periodic point set $S$ is given by a pair (cell $U$, motif $M$).
Here $U$ is a \emph{unit cell} (parallelepiped) spanned by a linear basis $v_1,\dots, v_n$ of $\R^n$, which generates the \emph{lattice} $\La=\{\sum\limits_{i=1}^n c_i v_i \vl c_i\in\Z \}$.
A \emph{periodic point set} $S=M+\La$ is obtained by shifting a finite \emph{motif} $M\subset U$ of points along all vectors $v\in\La$.
Fig.~\ref{fig:crystal_invariants} illustrates the problem of transforming ambiguous input into invariants that can distinguish periodic sets up to isometry. 

Arguably the simplest isometry invariant of a crystal is its density $\rho$.
Without distinguishing atoms in a periodic point set $S$, the density $\rho(S)$ is the number $m$ of points in a unit cell $U$, divided by the cell volume $\vol[U]$.
The density $\rho$ distinguishes hexagonal and square lattices in Fig.~\ref{fig:crystal_invariants} but is insensitive to perturbations shown in Fig.~\ref{fig:perturbations}.
Though many real crystals are dense and can not be well-separated by density, energy landscapes are still visualized as energy-vs-density plots in Fig.~\ref{fig:T2CSP}.   
The single-value density $\rho$ has been recently extended to the sequence of density functions $\psi_k(t)$ \cite{edels2021}.
For any integer $k\geq 1$, the \emph{density function} $\psi_k(t)$ measures the volume of the regions within a unit cell $U$ covered by $k$ balls with radius $t\geq 0$ and centers at all points $p\in M$, divided by $\vol[U]$.
\medskip

Though these isometry invariants have helped to identify a missing crystal in the Cambridge Structural Database, their running time cubically depends on $k$, which is a bit slow for big datasets.
The following invariants are much faster.
\medskip

\begin{figure}[h]
\includegraphics[height=45mm]{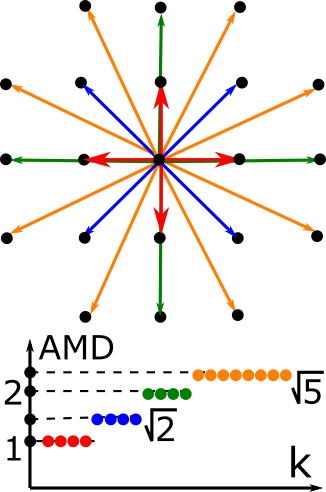}
\hspace*{2mm}
\includegraphics[height=45mm]{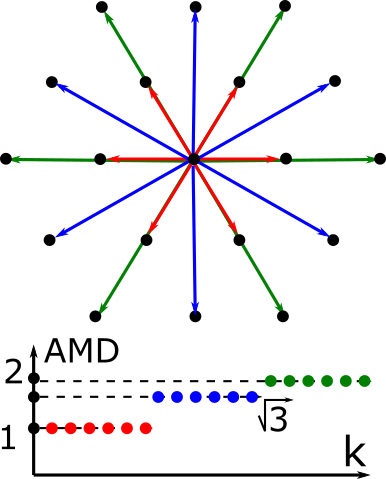}
\hspace*{2mm}
\includegraphics[height=45mm]{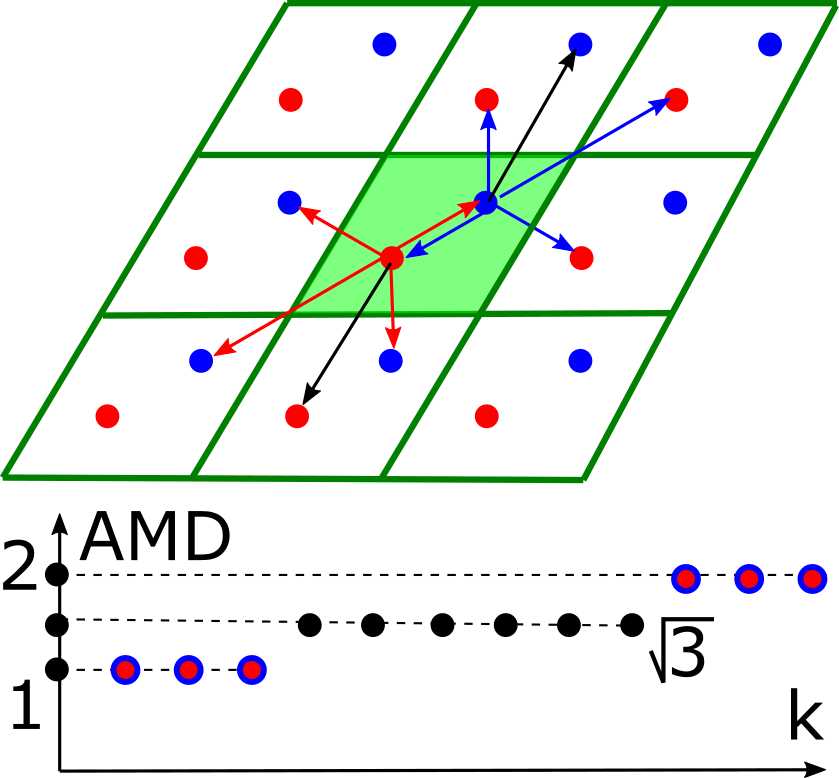}
\caption{\cite[Fig.~4]{widdowson2022average} \textbf{Left}: in the square lattice, the $k$-th neighbors of the origin and corresponding $\AMD_k$  are shown in the same color, e.g. the shortest axis-aligned distances $\AMD_1=\dots=\AMD_4=1$ are in red, the longer diagonal distances $\AMD_5=\dots=\AMD_8=\sqrt{2}$ are in blue.
\textbf{Middle}: in the hexagonal lattice, the shortest distances are in red: $\AMD_1=\dots=\AMD_6=1$.
\textbf{Right}: AMD for a non-lattice set.}
\label{fig:AMD}
\end{figure}

Let a periodic point set $S\subset\R^n$ have points $p_1,\dots,p_m$ in a unit cell.
For any $k\geq 1$ and $i=1,\dots,m$, the $i$-th row of the $m\times k$ matrix $D(S;k)$ consists of the ordered distances $d_{i1}\leq\cdots\leq d_{ik}$ measured from the point $p_i$ to its first $k$ nearest neighbors within the infinite set $S$, see Fig.~\ref{fig:AMD}.
The \emph{Average Minimum Distance}
$\AMD_k(S)=\dfrac{1}{m}\sum\limits_{i=1}^m d_{ik}$ is the average of the $k$-th column in $D(S;k)$.
\bs

\vspace*{-6mm}
\vspace*{-4mm}

\cite[Theorem~4]{widdowson2022average} proves that AMD is an isometry invariant independent of a unit cell.
The AMD invariants are similar to radial distribution functions \cite{schutt14represent} and related density-based invariants \cite{valle2010crystal}.
The AMD definition has no manually chosen thresholds such as cut-off radii or tolerances.
The length $k$ of the vector $\AMD^{(k)}=(\AMD_1,\dots,\AMD_k)$ is not a parameter in the sense that increasing $k$ only adds new values without changing previous ones.
Hence $k$ can be considered as an order of approximation, similarly to an initial length of a DNA code.
\medskip

We have no examples of non-isometric sets that have identical infinite AMD sequences.
Hence AMD can be complete at least for periodic sets in general position so that if two sets $S,Q$ have $\AMD(S)=\AMD(T)$, then $S,Q$ are isometric.
More recently, the isometry classification of all periodic point sets was reduced to an \emph{isoset} \cite{anosova2021isometry}, which is a collection of atomic environments considered modulo rotations and up to a \emph{stable} radius $\al$.
This stable radius is defined for a given crystal and any two crystals can be compared by isosets of their maximum radius so that two sets $S,Q$ are isometric if and only if their isosets are equivalent.
\medskip

This paper uses AMD invariants due to their easy interpretability and fast running time.
$\AMD_k(S)$ asymptotically approaches $c(S)\sqrt[n]{k}$, where $c(S)$ is related to the density of a periodic point set $S\subset\R^n$, see \cite[Theorem~13]{widdowson2022average}.
A near linear computational time \cite[Theorem~14]{widdowson2022average} of $\AMD_k$ in both $m,k$ translates into milliseconds on a modest laptop, which allowed us to visualize all 229K organic molecular crystals from the Cambridge Structural Database in a few hours.  

\section{Continuity of the energy in terms of AMD invariants}
\label{sec:continuous_energy}

To express continuity of AMD and other invariants under perturbations, we use the maximum displacement of atoms formalized by the \emph{bottleneck distance} $d_B$ as follows.
For any bijection $g:S\to Q$ between periodic point sets, the maximum displacement is $d_g(S,Q)=\sup\limits_{p\in S}|g(p)-p|$.
After minimizing over all bijections $g:S\to Q$, we get the \emph{bottleneck distance} $d_B(S,Q)=\inf\limits_{g:S\to Q} d_g(S,Q)$.
\medskip

\noindent
\textbf{The structure-property hypothesis} says that all properties of a crystal should be determined by its geometric structure.
Understanding how any property can be explicitly computed from a crystal structure would replace trial-and-error methods by a guided discovery to find crystals with desired properties.
\medskip

Most current attempts are based on black-box machine learning of properties from crystal descriptors, not all of which are invariants up to isometry.
All machine learning tools rely on the usually implicit assumption that small perturbations in input data lead to relatively small perturbations in outputs. 
\medskip

\noindent
\textbf{Continuity of a structure-property relation} can be mathematically expressed as Lipschitz continuity \cite[section 9.4]{osearcoid2006metric}:  $|E(S)-E(Q)|\leq \la d(S,Q)$, where $\la$ is a constant, $E$ is a crystal property such as the lattice energy, $d(S,Q)$ is a distance satisfying all metric axioms on crystals $S,Q$ or their invariants. 
The above inequality should hold for all crystals $S,Q$ with small distances $d(S,Q)<\de$, where a threshold $\de$ may depend on a property $E$ or a metric $d$, not on $S,Q$. 
\medskip

The continuity above sounds plausible and seems necessary for the structure-property hypothesis.
Indeed, if even small perturbations of a geometric structure drastically change crystal properties, then any inevitably noisy structure determination would not suffice to guarantee desired properties of a crystal.
\medskip

\begin{figure}[h!]
\includegraphics[width=\textwidth]{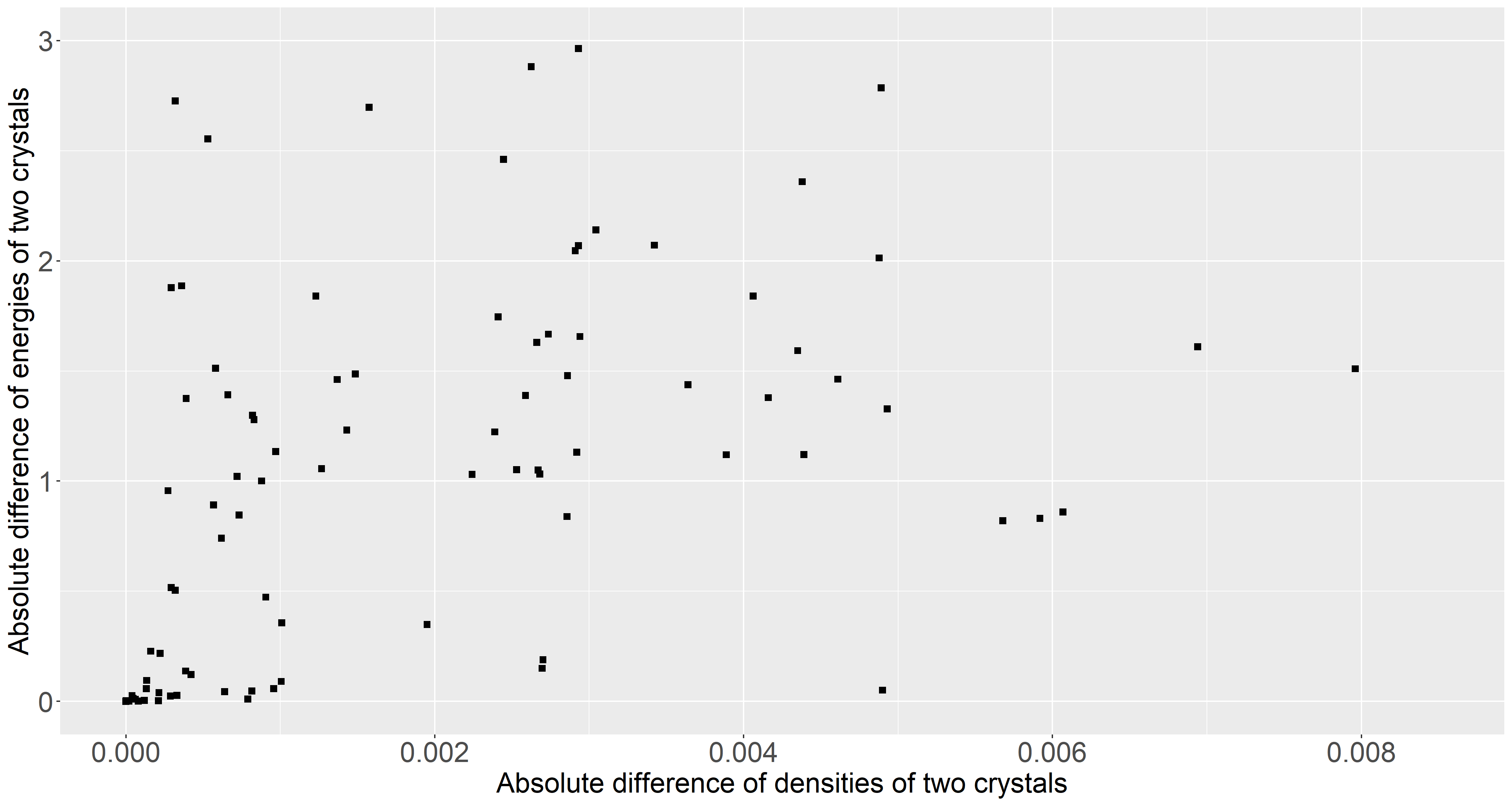}
\caption{5679 crystals in Fig.~\ref{fig:T2CSP} have the density range $[0.3,1.4]$.
Many crystals have differences in densities within $0.003$g/cm$^3$ and differences in the energy up to 3kJ/mole.}
\label{fig:energy_dif-vs-density}
\end{figure}

\begin{figure}[h!]
\includegraphics[width=\textwidth]{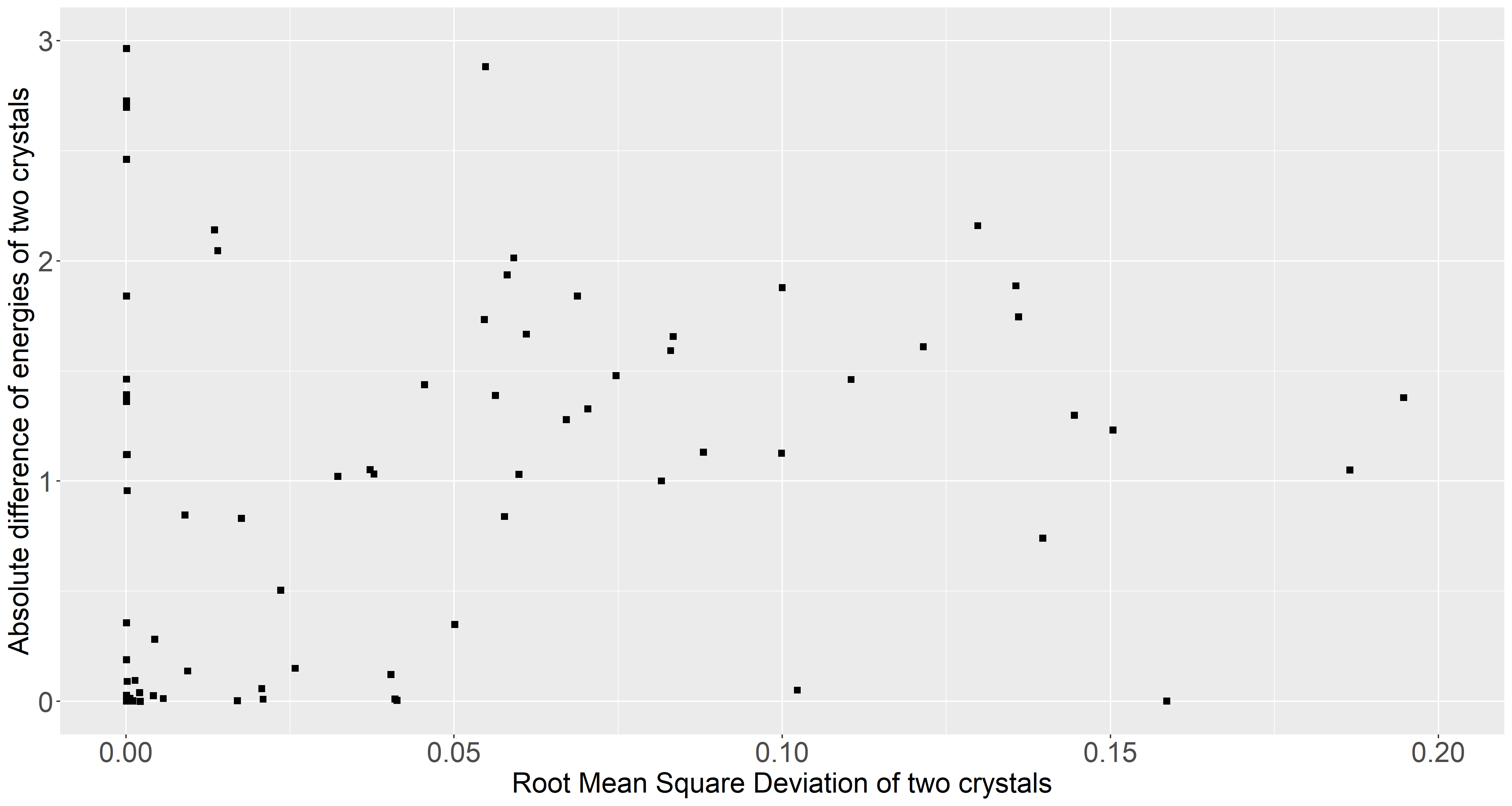}
\caption{Crystal pairs with RMSD $<0.1\AA$ have energy differences up to 3kJ/mole.}
\label{fig:energy_dif-vs-RMSD}
\end{figure}

\begin{figure}[h!]
\includegraphics[width=\textwidth]{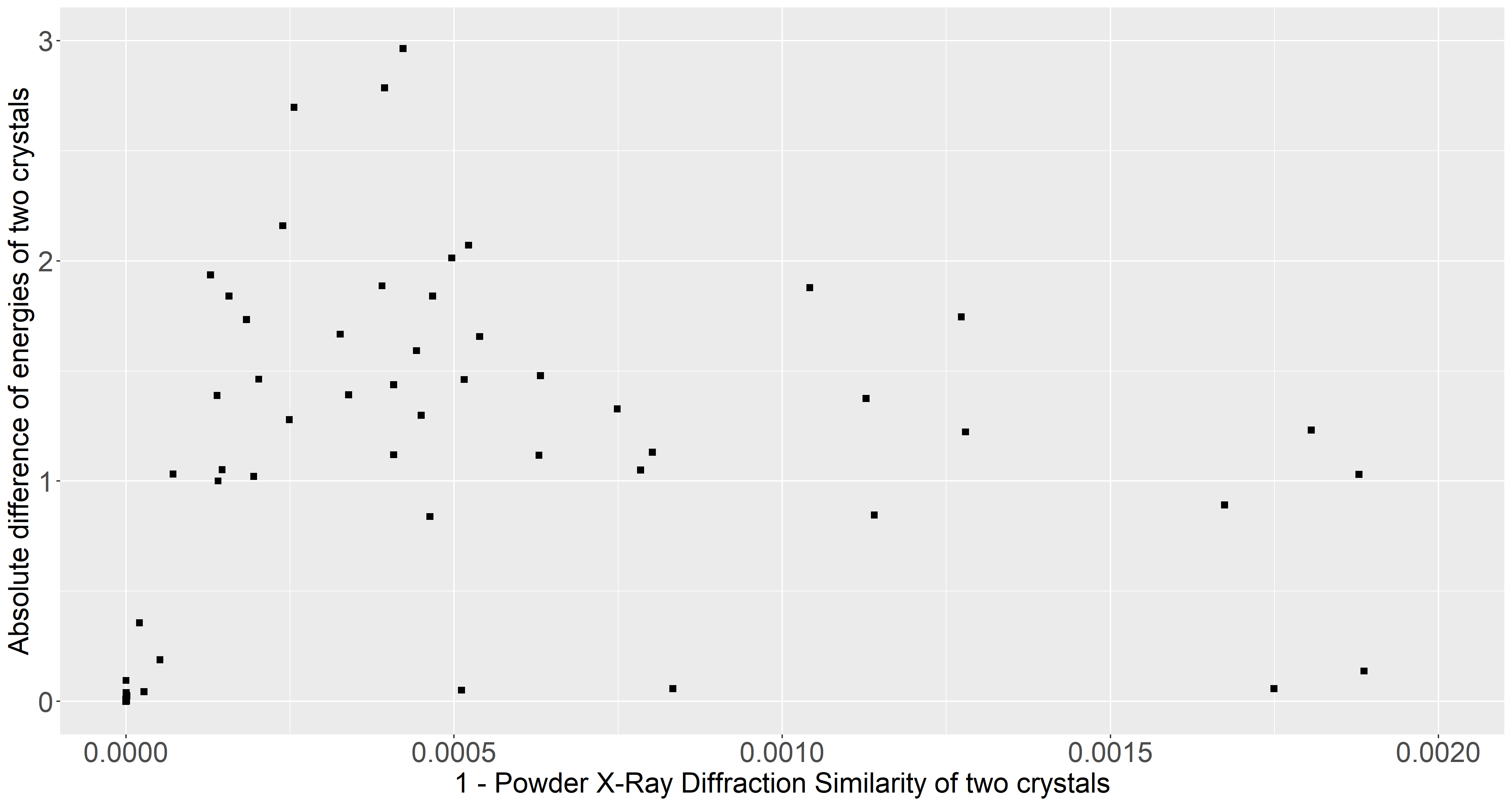}
\caption{Crystal pairs with PRXD similarity $>0.9995$ have big energy differences.}
\label{fig:energy_dif-vs-PXRD}
\end{figure}

Fig.~\ref{fig:energy_dif-vs-density},\ref{fig:energy_dif-vs-RMSD},\ref{fig:energy_dif-vs-PXRD} show that the past methods of characterizing crystal similarity are insufficient to guarantee the above continuity of the lattice energy. 
These results were obtained on the T2 dataset of 5679 simulated crystals reported in \cite{pulido2017functional}.
Each square dot represents a pair of crystals with differences in past descriptors on the horizontal axis and differences in energies on the vertical axis.
\medskip

Fig.~\ref{fig:energy_dif-vs-density} shows dozens of crystal pairs with very close densities and rather different lattice energies, which means that the energy discontinuously varies relative to the density.
This failure of a single-value descriptors might not be surprising not only for crystals, which are often very dense materials, but also for other real-life scenarios.
For example, many people have the same height and very different weights.
However, the density is still used to represent a crystal structure in CSP landscapes such as Fig.~\ref{fig:T2CSP}.
Indeed, the density is an isometry invariant, which is continuous (actually, constant) under perturbations, see Fig.~\ref{fig:perturbations}.
\medskip

Fig.~\ref{fig:energy_dif-vs-RMSD} illustrates a similar conclusion for the traditional packing similarity measured by the COMPACK algorithm \cite{chisholm2005compack} as the Root Mean Square Deviation (RMSD) of atomic positions matched between up to 15 (by default) molecules in two crystals.
This similarity relies on two extra thresholds for atomic distances and angles whose values affect the RMSD.
For example, when only one of 15 molecules is matched, the RMSD is exactly 0, because all 5679 crystals are based on the same T2 molecule in Fig.~\ref{fig:T2CSP}.
Nonetheless, this packing similarity can visually confirm that nearly identical crystals nicely overlap each other.
\medskip

\begin{figure}[h!]
\includegraphics[width=\textwidth]{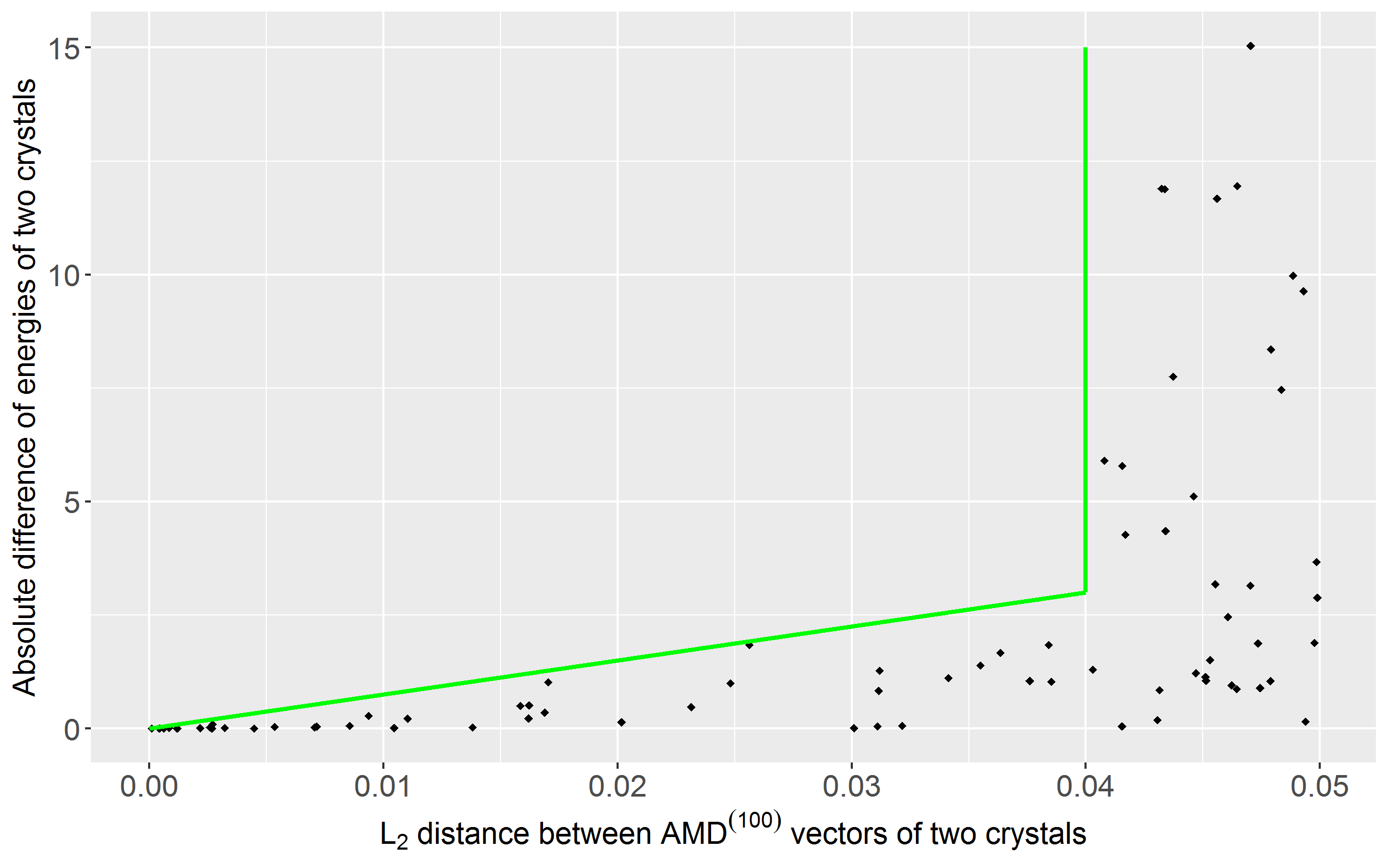}
\caption{The green line $|\De E|=75L_2$ over $L_2\in[0,0.04]$ shows that if crystals have a distance $L_2<0.04\AA$ between $\AMD^{(100)}$ vectors, their energies differ by at most $75L_2$.}
\label{fig:energy_dif-vs-L2_AMD100}
\end{figure}

The powder X-ray diffraction (PXRD) similarity has the range [0,1] with values close to 1 indicating closeness of diffraction patterns.
Fig.~\ref{fig:energy_dif-vs-PXRD} has $1-$PXRD on the horizontal axis and similarly to the above two plots shows many pairs of nearly identical crystals (with PXRD above 0.9995) with rather different energies.
Despite Fig.~\ref{fig:energy_dif-vs-density},\ref{fig:energy_dif-vs-RMSD},\ref{fig:energy_dif-vs-PXRD} illustrating the discontinuity of the lattice energy with respect to traditional similarity measures of crystals, we should not despair.
\medskip

The new AMD invariants detect tiny differences in crystal structures and are continuous under perturbations in the bottleneck distance \cite[Theorem~9]{widdowson2022average}: $|\AMD_k(S)-\AMD_k(Q)|\leq 2d_B(S,Q)$ if the bottleneck distance $d_B$ is less than half of the minimum distance between points in any of periodic sets $S,Q\subset\R^n$. 
\medskip

\begin{figure}[h!]
\includegraphics[width=\textwidth]{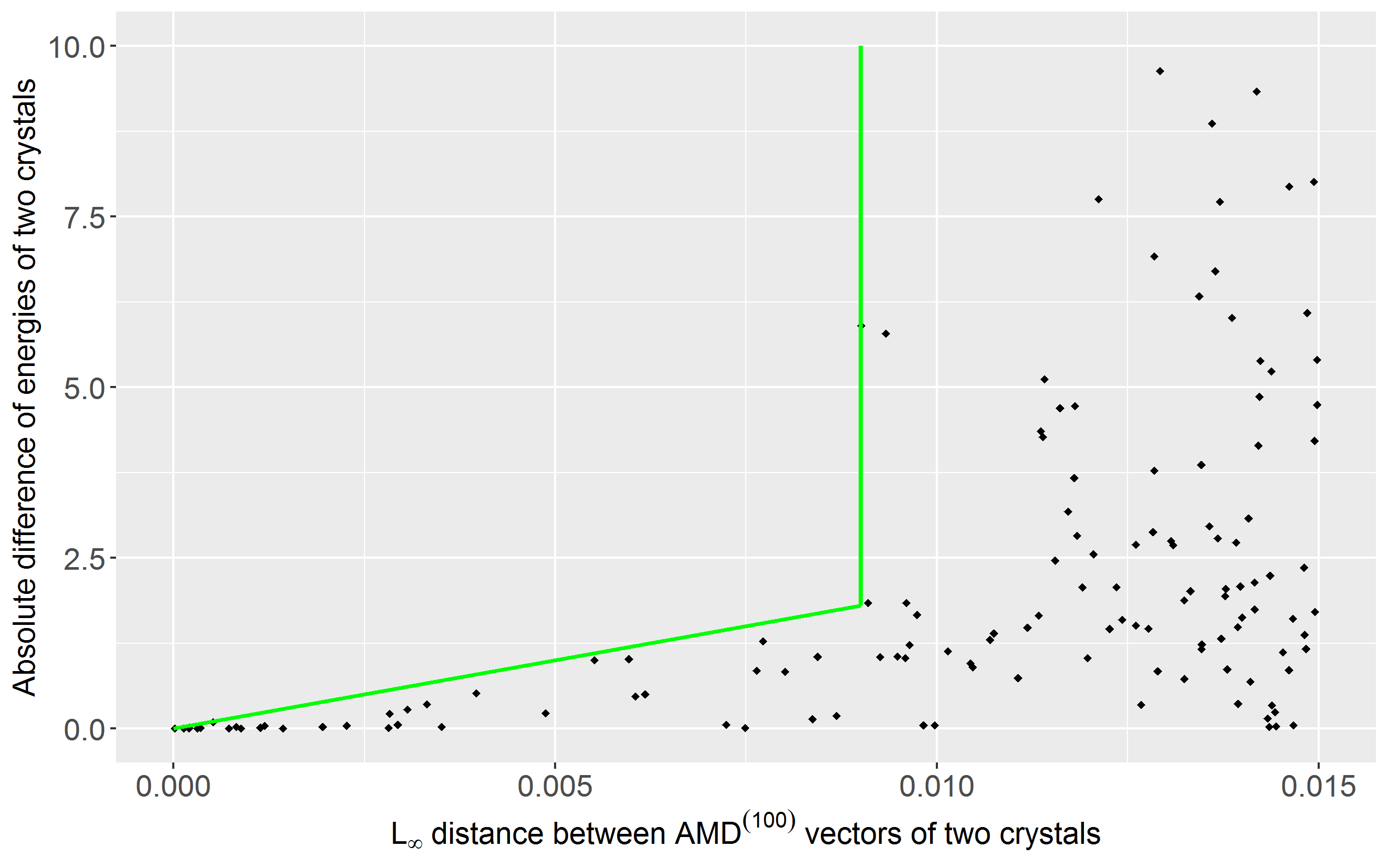}
\caption{The green line $|\De E|=200L_{\infty}$ over $[0,0.009]$ shows that if crystals have a distance $L_{\infty}<0.009\AA$ between $\AMD^{(100)}$ vectors, their energies differ by at most $200L_{\infty}$.}
\label{fig:energy_dif-vs-Linf_AMD100}
\end{figure}

\begin{figure}[h!]
\includegraphics[width=\textwidth]{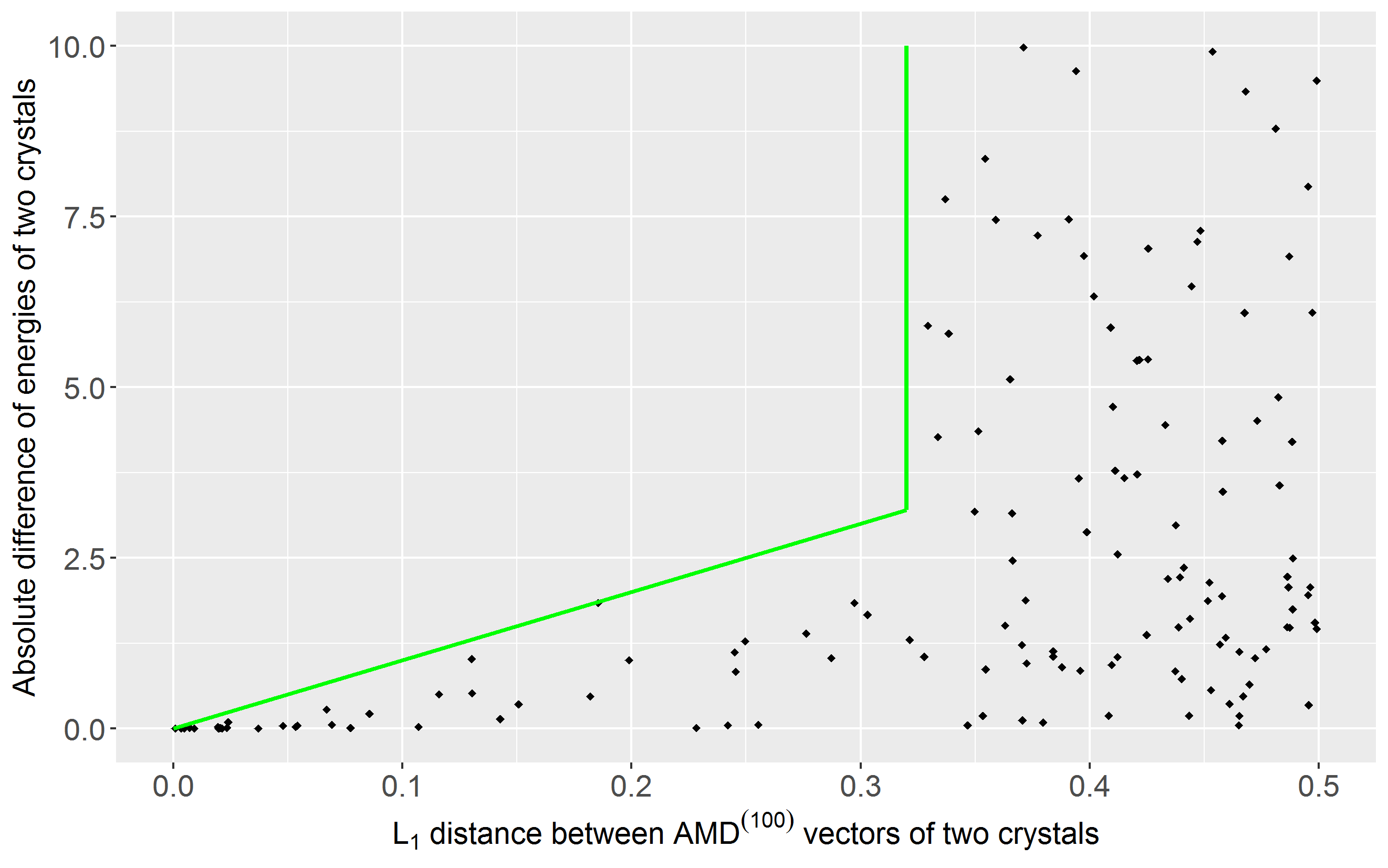}
\caption{The green line $|\De E|=10L_1$ over $L_1\in[0,0.32]$ shows that if crystals have a distance $L_1<0.32\AA$ between $\AMD^{(100)}$ vectors, their energies differ by at most $10L_1$.}
\label{fig:energy_dif-vs-L1_AMD100}
\end{figure}

Even more importantly, Fig.~\ref{fig:energy_dif-vs-L2_AMD100}, \ref{fig:energy_dif-vs-Linf_AMD100},\ref{fig:energy_dif-vs-L1_AMD100}  show that the lattice energy continuously changes with respect to AMD invariants on the same T2 dataset.
Each rhombic dot in Fig.~\ref{fig:energy_dif-vs-L2_AMD100},\ref{fig:energy_dif-vs-Linf_AMD100},\ref{fig:energy_dif-vs-L1_AMD100}  represents one pairwise comparison between $\AMD^{(100)}$ vectors of length $k=100$ for two T2 crystals.
The distances between vectors $p=(p_1,\dots,p_k)$ and $q=(q_1,\dots,q_k)$ on the horizontal axis are computed by 
the Euclidean metric $L_2(p,q)=\sqrt{\sum\limits_{i=1}^k|p_i-q_i|^2}$,
the Chebyshev metric $L_{\infty}(p,q)=\max\limits_{i=1,\dots,k}|p_i-q_i|$ and the Manhattan metric $L_1(p,q)=\sum\limits_{i=1}^k|p_i-q_i|$.
\medskip

Despite the T2 dataset being thoroughly filtered out by several tools to remove near duplicates, Fig.~\ref{fig:energy_dif-vs-L2_AMD100}, \ref{fig:energy_dif-vs-Linf_AMD100} include several pairs whose AMD invariants are very close, though not identical.
In all these cases the corresponding crystals also have very close energies, which can be quantified via Lipschitz continuity as follows.
\medskip

In Fig.~\ref{fig:energy_dif-vs-L2_AMD100} the Lipschitz continuity for the energy $|\De E|=|E(S)-E(Q)|\leq \la_{2}L_2(\AMD^{(100)}(S),\AMD^{(100)}(Q))$ holds for $\la_2=200$ and all pairs of crystals $S,Q$ whose $\AMD^{(100)}$ vectors have a Euclidean distance $L_2<\de_2=0.04\AA$.
\medskip

Visually, all these pairs are below the green line $\De E=200L_2$ up to the distance threshold $\de_2=0.04\AA$.
If distances between crystals become too large, a single-value metric cannot guarantee close values of energy.
Using the geographic analogy, the further we travel from any fixed location on planet Earth, the more variation in physical properties such as the altitude we should expect. 
\medskip

Fig.~\ref{fig:energy_dif-vs-Linf_AMD100} similarly illustrates continuity of the lattice energy with respect to the metric 
$L_{\infty}(p,q)=\max\limits_{i=1,\dots,k}|p_i-q_i|$
between $\AMD^{(100)}$ vectors.
All pairs of crystals with distances $L_{\infty}<\de_{\infty}=0.009\AA$ have energy differences less than $\la_{\infty}L_{\infty}$ with $\la_{\infty}=200$, so all corresponding dots are below the green line $|\De E|=200 L_{\infty}$.
\medskip

Fig.~\ref{fig:energy_dif-vs-L1_AMD100} shows that the lattice energy continuously behaves for the metric 
$L_1(p,q)=\sum\limits_{i=1}^k|p_i-q_i|$ between $\AMD^{(100)}$ vectors.
All pairs of crystals with distances $L_1<\de_1=0.32\AA$ have energy differences less than $\la_1 L_1$ with $\la_1=10$, so all corresponding dots are below the green line $|\De E|=10 L_1$.
\medskip

The thresholds $\de_1=0.32$ and $\de_2=0.04$ are larger than $\de_{\infty}=0.009\AA$, because the metrics $L_1,L_2$ sum up all deviations between corresponding coordinates of $\AMD^{(100)}$ vectors, while the metric $L_{\infty}$ measures only the maximum deviation.
\medskip

If we tried to fit Lipschitz continuity for the past descriptors (density, RMSD, PXRD) in Fig.~\ref{fig:energy_dif-vs-density},\ref{fig:energy_dif-vs-RMSD},\ref{fig:energy_dif-vs-PXRD} similarly to AMD invariants above, corresponding green lines would be almost vertical with huge slopes or gradients (Lipschitz constants).

\section{Fast predictions of the energy by AMD invariants}
\label{sec:energy_predictions}

This section describes the second important contribution by showing that continuity of AMD from section~\ref{sec:continuous_energy} leads to state-of-the-art energy predictions. 
\medskip

\noindent
\textbf{The energy prediction problem} is to infer the lattice energy from a crystal structure, for example by using a dataset of ground truth energies for training.
\medskip

The descriptors in Fig.~\ref{fig:energy_dif-vs-density},\ref{fig:energy_dif-vs-RMSD},\ref{fig:energy_dif-vs-PXRD} cannot be justifiably used to resolve the above problem because of their discontinuity.
Indeed, if we input a slightly different (say, experimental) crystal, we expect a close value of energy in the output.
\medskip

\begin{figure}[h!]
\includegraphics[height=40mm]{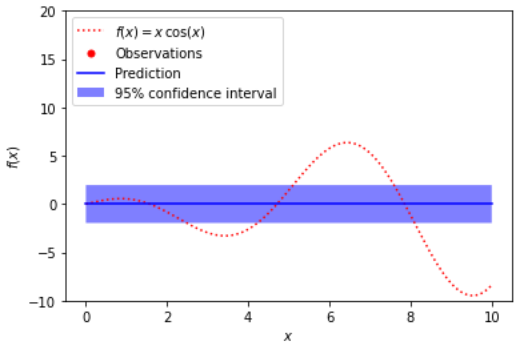}
\includegraphics[height=40mm]{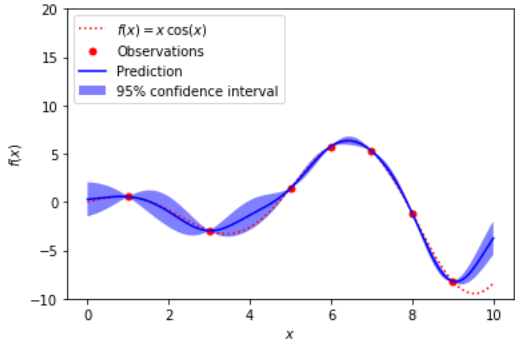}
\caption{Gaussian Process tries to predict values of $f(x)=x\cos x$ by training on observed data points.
\textbf{Left}: an initial prediction is 0 for any $x$.
\textbf{Right}: predictions substantially improve after training on six data points under the natural assumption that the underlying function is \emph{continuous}, so continuity is important for learning.}
\label{fig:Gaussian_process_illustration}
\end{figure}

First we describe the Gaussian Process Regression \cite{ki2006gaussian} as implemented in SciKit Learn \cite{pedregosa2011scikit}, see Fig.~\ref{fig:Gaussian_process_illustration}, which achieved the best results on the T2 dataset of 5679 crystals.
Initially each T2 crystal is converted into a periodic point set $S$ by placing a zero-sized point at every atomic center.
Then each $S$ is represented by its $\AMD^{(k)}(S)$ vector of a fixed length in the range $k=50,100,\dots,500$.
The base distance $d$ between $\AMD^{(k)}$ vectors was chosen as $L_{\infty}$ due to the smallest Lipschitz constant $\la=2$ in the continuity property $|\AMD_k(S)-\AMD_k(Q)|\leq \la d_B(S,Q)$.
For the metrics $L_1,L_2$, the Lipschitz constants would be $2k,2\sqrt{k}$. 
\medskip

For any pair of crystals $S,Q$, we consider the Rational Quadratic Kernel
 $K(S,Q)=\left(1+\dfrac{d^2(S,Q)}{2\al l^2}\right)^{-\al}$, where $\al,l$ are scale parameters optimized by training.
For a single prediction run, the whole T2 dataset was randomly split into 80\% training subset and remaining 20\% test subset of $m=1136$ crystals.
\medskip

Table~\ref{tab:predictions} shows three types of errors, each averaged over 10 runs above: 
$\RMSE=\sqrt{\dfrac{1}{m}\sum\limits_{i=1}^m |E_{true}(S_i)-E_{pred}(S_i)|^2}$ is the root mean square error in the lattice energy averaged over $m$ crystals $S_1,\dots,S_m$ from the test subset, then
$\MAE =\dfrac{1}{m}\max\limits_{i=1,\dots,m} |E_{true}(S_i)-E_{pred}(S_i)|$ is the mean absolute error and the mean absolute percentage error
$\MAPE=\dfrac{1}{m}\max\limits_{i=1,\dots,m} \dfrac{|E_{true}(S_i)-E_{pred}(S_i)|}{E_{true}(S_i)}$.
Each value has the empirical standard deviation $\pm\std$ computed over 10 runs.
\medskip

\begin{table}[h]
\caption{The Gaussian Process with the Rational Quadratic Kernel predicts the energy reported in \cite{pulido2017functional} with the mean absolute error (MAE) of less than 5kJ/mole on $m=1136$ crystals by training on the isometry invariants $\AMD^{(k)}$ of 4543 crystals for various $k$.}
\label{tab:predictions}
\begin{tabular}{c|p{20mm}|p{20mm}|p{20mm}|c|c}
$k$ & $\RMSE\pm\std$ & $\MAE\pm\std$ & $\MAPE\pm\std$ & training time, sec & full test time, ms\\
\hline
50 & $6.503 \pm 0.123$ & $4.900\pm 0.86$ & $3.509\pm 0.059$ & $627\pm 85$ & $15961\pm 183$ \\
\hline
100 & $6.344\pm 0.152$ & {\bfseries $4.801\pm 0.103$} & $3.439\pm 0.070$ & $349\pm 47$ & $7979\pm 564$ \\
\hline
150 & $6.607\pm 0.119$ & $4.977\pm 0.077$ & $3.559\pm 0.053$ & $400\pm 23$ & $12789\pm 203$ \\
\hline
200 & $6.617\pm 0.147$ & $4.966\pm 0.114$ & $3.554\pm 0.079$ & $506\pm 40$ & $15943\pm 46$ \\
\hline 
250 & $6.517\pm 0.109$ & $4.914\pm 0.082$ & $3.514\pm 0.055$ &
$574\pm 91$ & $16464\pm 193$ \\
\hline
300 & $6.632\pm 0.139$ & $5.003\pm 0.092$ & $3.577\pm 0.062$ & $545\pm 15$ & $16431\pm 52$ \\
\hline
350 & $6.615\pm 0.077$ & $4.990\pm 0.077$ & $3.581\pm 0.053$ & $500\pm 22$ & $12395\pm 44$ \\
\hline
400 & $6.611\pm 0.149$ & $4.984\pm 0.080$ & $3.569\pm 0.053$ & $585\pm 25$ & $17906\pm 201$ \\
\hline
450 & $6.559\pm 0.179$ & $4.954\pm 0.127$ & $3.545\pm 0.085$ & $512\pm 21$ & $12927\pm 67$ \\
\hline
500 & $6.622\pm 0.116$ & $5.004\pm 0.092$ & $3.581\pm 0.068$ & $598\pm 24$ & $18429\pm 219$
\end{tabular}
\end{table}

Table~\ref{tab:predictions} shows that the errors $\RMSE, \MAE, \MAPE$ are consistent across different values of $k$.
The key advantage over past methods is the speed: less than 10 min for training for 4600 vectors $\AMD^{(k)}$ on Intel Xeon CPU at 2.3 GHz. 
The last column shows the full test time on $m=1136$ crystals, so the average time per crystal is more than 1000 times faster.
The smallest mean absolute error $\MAE\approx 4.8$kJ/mole corresponds to about 7.4 milliseconds (ms) per crystal.
\medskip

The computation of $\AMD^{(k)}$ asymptotically has a near linear time in $k$ and the number of atoms in a unit cell by \cite[Theorem~14]{widdowson2022average}, which needs only 27ms on average per T2 crystal for $k=1000$ on a similar desktop.
This ultra-fast speed allowed us to visualize for the first time all 229K molecular organic crystals from the Cambridge Structural Database in less than 9 hours, see \cite[appendix~D]{widdowson2022average}. 
\medskip

We have tried other types of kernels: the matern and linear kernels gave slightly larger errors, the squared exponential was worse for some $k$.
We also considered another version of the T2 dataset without hydrogens (32 atoms per molecule instead of 46), which gave a bit bigger error for all kernels above.
\medskip

Instead of AMD invariants, we trained the Gaussian Process Regression on the density functions  $\psi_k(t)$ \cite{edels2021}, which are continuous isometry invariants extending the single-value density for a variable radius $t\geq 0$.
The average errors of AMD-based predictions were smaller than for the density functions $\psi_k$, which are also slower to compute than AMD, asymptotically in a cubic time in $k$.
\medskip

Finally, the Random Forest \cite{myles2004introduction} and Dense Neural Network \cite{goodfellow2016deep} trained on AMD and density functions performed slight worse than the Gaussian Process, though the training and text times were much faster (seconds instead of minutes).
The experiments above are reported in the dissertation of the first author \cite{ropers2021applying}.
 
\section{Conclusions and a discussion of future developments}
\label{sec:discussion}

This paper has demonstrated that the recently developed continuous isometry invariants can provide insights undetected by traditional similarity measures.
\medskip

In section~\ref{sec:continuous_energy} Fig.~\ref{fig:energy_dif-vs-density},\ref{fig:energy_dif-vs-RMSD},\ref{fig:energy_dif-vs-PXRD} show that many crystals can have almost identical density, RMSD, PXRD patterns but rather different lattice energies. 
On the same T2 dataset \cite{pulido2017functional} Fig.~\ref{fig:energy_dif-vs-L2_AMD100},\ref{fig:energy_dif-vs-Linf_AMD100},\ref{fig:energy_dif-vs-L1_AMD100} show that the lattice energy satisfies the Lipschitz continuity $|E(S)-E(Q)|\leq\la d(S,Q)$ for a fixed constant $\la$ and all crystals $S,Q$ whose AMD invariants are close with respect to the metrics $L_1,L_2,L_{\infty}$.
\medskip

In section~\ref{sec:energy_predictions} the standard kernel methods trained only on 100 isometry invariants $\AMD^{(100)}$ achieved the state-of-the-art mean absolute error of less than 5kJ/mole in energy.
The key achievement is the speed of training (about 10 min for 4543 crystals on a modest desktop) and testing, which run in milliseconds per crystal.
The code of experiments in section~\ref{sec:energy_predictions} is available on GitHub \cite{ropers2021applying}.
\medskip

It should not be surprising that the lattice energy can be efficiently predicted from distance-based invariants without any chemical information.
Indeed, if one atom is replaced by a different chemical element, then inter-atomic distances to neighbors inevitably change, even if slightly.
These differences in distances can be detected, also after averaging over all motif points.
So AMD invariants should pick up differences in crystals after swapping chemically different atoms. 
\medskip

We don't know any non-isometric periodic point sets that have identical infinite sequences $\{\AMD_k\}_{k=1}^{+\infty}$.
Claiming such a counter-example to completeness requires a theoretical proof, because any computation outputs AMD values only up to a finite $k$.
Future work will discuss the stronger Pointwise Distance Distributions, which avoid averaging over motif points, but keep the isometry invariance and continuity under perturbations of points for a suitable metric.

%
%
%
\bibliographystyle{splncs04}
\bibliography{energy_predictions}

\begin{thebibliography}{10}
\providecommand{\url}[1]{\texttt{#1}}
\providecommand{\urlprefix}{URL }
\providecommand{\doi}[1]{https://doi.org/#1}

\bibitem{anosova2021introduction}
Anosova, O., Kurlin, V.: Introduction to periodic geometry and topology.
  arXiv:2103.02749  (2021)

\bibitem{anosova2021isometry}
Anosova, O., Kurlin, V.: An isometry classification of periodic point sets. In:
  Proceedings of Discrete Geometry and Mathematical Morphology (2021)

\bibitem{behler2011atom}
Behler, J.: Atom-centered symmetry functions for constructing high-dimensional
  neural network potentials. The Journal of chemical physics  \textbf{134}(7),
  074106 (2011)

\bibitem{chisholm2005compack}
Chisholm, J., Motherwell, S.: Compack: a program for identifying crystal
  structure similarity using distances. J. Applied Crystallography
  \textbf{38}(1),  228--231 (2005)

\bibitem{edels2021}
Edelsbrunner, H., Heiss, T., Kurlin, V., Smith, P., Wintraecken, M.: The
  density fingerprint of a periodic point set. In: Proceedings of SoCG (2021)

\bibitem{egorova2020multifidelity}
Egorova, O., Hafizi, R., Woods, D.C., Day, G.M.: Multifidelity statistical
  machine learning for molecular crystal structure prediction. The Journal of
  Physical Chemistry A  \textbf{124}(39),  8065--8078 (2020)

\bibitem{goodfellow2016deep}
Goodfellow, I., Bengio, Y., Courville, A.: Deep learning, vol.~1. MIT Press
  (2016)

\bibitem{gross2013density}
Gross, E., Dreizler, R.: Density functional theory, vol.~337. Springer Science
  \& Business Media (2013)

\bibitem{ki2006gaussian}
KI~Williams, C.: Gaussian processes for machine learning. Taylor \& Francis
  (2006)

\bibitem{mosca2020voronoi}
Mosca, M., Kurlin, V.: Voronoi-based similarity distances between arbitrary
  crystal lattices. Crystal Research and Technology  \textbf{55}(5),  1900197
  (2020)

\bibitem{myles2004introduction}
Myles, A.J., Feudale, R.N., Liu, Y., Woody, N.A., Brown, S.D.: An introduction
  to decision tree modeling. Journal of Chemometrics  \textbf{18}(6),  275--285
  (2004)

\bibitem{niketic2012consistent}
Niketic, S.R., Rasmussen, K.: The consistent force field: a documentation,
  vol.~3. Springer Science \& Business Media (2012)

\bibitem{oganov2011modern}
Oganov, A.: Modern methods of crystal structure prediction. Wiley \& Sons
  (2011)

\bibitem{osearcoid2006metric}
O'Searcoid, M.: Metric spaces. Springer Science \& Business Media (2006)

\bibitem{pedregosa2011scikit}
Pedregosa, F., Varoquaux, G., Gramfort, A., Michel, V., Thirion, B., Grisel,
  O., Blondel, M., Prettenhofer, P., Weiss, R., Dubourg, V., et~al.:
  Scikit-learn: Machine learning in python. Journal of machine Learning
  research  \textbf{12},  2825--2830 (2011)

\bibitem{pulido2017functional}
Pulido, A., Chen, L., Kaczorowski, T., Holden, D., Little, M., Chong, S.,
  Slater, B., McMahon, D., Bonillo, B., Stackhouse, C., Stephenson, A., Kane,
  C., Clowes, R., Hasell, T., Cooper, A., Day, G.: Functional materials
  discovery using energy--structure maps. Nature  \textbf{543},  657--664
  (2017)

\bibitem{ropers2021applying}
Ropers, J.: Applying machine learning to geometric invariants of crystals
  (2021), \url{https://github.com/JRopes/CrystalEnergyPrediction}

\bibitem{sacchi2020same}
Sacchi, P., Lusi, M., Cruz-Cabeza, A.J., Nauha, E., Bernstein, J.: Same or
  different--that is the question: identification of crystal forms from crystal
  structure data. CrystEngComm  \textbf{22}(43),  7170--7185 (2020)

\bibitem{schutt14represent}
Sch{\"u}tt, K., Glawe, H., Brockherde, F., Sanna, A., M{\"u}ller, K.R., Gross,
  E.: How to represent crystal structures for machine learning: Towards fast
  prediction of electronic properties. Physical Review B  \textbf{89}(20),
  205118 (2014)

\bibitem{smith17ani}
Smith, J., Isayev, O., Roitberg, A.: An extensible neural network potential
  with dft accuracy at force field computational cost. Chem. Science
  \textbf{8},  3192--3203 (2017)

\bibitem{valle2010crystal}
Valle, M., Oganov, A.R.: Crystal fingerprint space--a novel paradigm for
  studying crystal-structure sets. Acta Crystallographica Section A:
  Foundations of Crystallography  \textbf{66}(5),  507--517 (2010)

\bibitem{wales2018exploring}
Wales, D.J.: Exploring energy landscapes. Annual review of physical chemistry
  \textbf{69},  401--425 (2018)

\bibitem{ward17including}
Ward, L., Liu, R., Krishna, A., Hegde, V., Agrawal, A., Choudhary, A.,
  Wolverton, C.: Including crystal structure attributes in machine learning
  models of formation energies via voronoi tessellations. Physical Review B
  \textbf{96}(2),  024104 (2017)

\bibitem{widdowson2021pointwise}
Widdowson, D., Kurlin, V.: Pointwise distance distributions of periodic sets,
  \url{https://arxiv.org/abs/2108.04798}

\bibitem{widdowson2022average}
Widdowson, D., Mosca, M., Pulido, A., Kurlin, V., Cooper, A.: Average minimum
  distances of periodic point sets. MATCH Communications in Mathematical and in
  Computer Chemistry, to appear  (2022), \url{https://arxiv.org/abs/2009.02488}

\end{thebibliography}

\end{document}